\documentclass[11pt,twoside]{article}

\usepackage{asp2014}

\aspSuppressVolSlug
\resetcounters

\begin{document}

\title{A Conception of Engineering Design for Remote Unattended Operation Public Observatory}

\author{Jun Han,$^1$ Dongwei Fan,$^1$ Chenzhou Cui,$^1$ Chuanzhong Wang,$^2$ Shanshan Li,$^1$ Linying Mi,$^1$ Zheng Li,$^1$ Yunfei Xu,$^1$ Boliang He,$^1$ Changhua Li,$^1$ Yihan Tao,$^1$ and Sisi Yang$^1$
\affil{$^1$National Astronomical Observatories, Chinese Academy of Science, Beijing, China; \email{hanjun@nao.cas.cn}}
\affil{$^2$Beijing University of Technology, Beijing, China} }

\paperauthor{Jun Han}{hanjun@nao.cas.cn}{}{National Astronomical Observatories, Chinese Academy of Science}{Center of Information and Computing}{Beijing}{Beijing}{100012}{China}
\paperauthor{Dongwei Fan}{fandongwei@nao.cas.cn}{}{National Astronomical Observatories, Chinese Academy of Science}{Center of Information and Computing}{Beijing}{Beijing}{100012}{China}
\paperauthor{Chenzhou Cui}{ccz@nao.cas.cn}{}{National Astronomical Observatories, Chinese Academy of Science}{Center of Information and Computing}{Beijing}{Beijing}{100012}{China}
\paperauthor{Chuanzhong Wang}{wangchuanzhong@nao.cas.cn}{}{Beijing University of Technology}{College of Mechanical Engineering and Applied Electronics Technology}{Beijing}{Beijing}{100022}{China}
\paperauthor{Shanshan Li}{lishanshan@nao.cas.cn}{}{National Astronomical Observatories, Chinese Academy of Science}{Center of Information and Computing}{Beijing}{Beijing}{100012}{China}
\paperauthor{Linying Mi}{mly@nao.cas.cn}{}{National Astronomical Observatories, Chinese Academy of Science}{Center of Information and Computing}{Beijing}{Beijing}{100012}{China}
\paperauthor{Zheng Li}{lizheng@nao.cas.cn}{}{National Astronomical Observatories, Chinese Academy of Science}{Center of Information and Computing}{Beijing}{Beijing}{100012}{China}
\paperauthor{Yunfei Xu}{xuyf@nao.cas.cn}{}{National Astronomical Observatories, Chinese Academy of Science}{Center of Information and Computing}{Beijing}{Beijing}{100012}{China}
\paperauthor{Boliang He}{hebl@nao.cas.cn}{}{National Astronomical Observatories, Chinese Academy of Science}{Center of Information and Computing}{Beijing}{Beijing}{100012}{China}
\paperauthor{Changhua Li}{lich@nao.cas.cn}{}{National Astronomical Observatories, Chinese Academy of Science}{Center of Information and Computing}{Beijing}{Beijing}{100012}{China}
\paperauthor{Yihan Tao}{y.tao@nao.cas.cn}{}{National Astronomical Observatories, Chinese Academy of Science}{Center of Information and Computing}{Beijing}{Beijing}{100012}{China}
\paperauthor{Sisi Yang}{yangss@nao.cas.cn}{}{National Astronomical Observatories, Chinese Academy of Science}{Center of Information and Computing}{Beijing}{Beijing}{100012}{China}

\begin{abstract}
Public observatory project is playing more and more important role in science popularization education and scientific research, and many amateur astronomers also have began to build their own observatories in remote areas. As a result of the limitation of technical condition and construction funds for amateur astronomers, their system often breaks down, and then a stable remote unattended operation system becomes very critical. Hardware connection and control is the basic and core part in observatory design. Here we propose a conception of engineering hardware design for public observatory operation as a bridge between observatory equipment and observation software. It can not only satisfy multiple observation mode requirement, but also save cost.
\end{abstract}


\section{Introduction}
Since the first microprocessor emerges, people have tried to make the telescope operation smarter and started numerous engineering attempts. Until the mid of 1960s, some began operating stably and the 8'' reflector telescope at the University of Wisconsin is one of earliest examples(\ \citet{1992ASPC...34....3C}). They could carry out astronomical photometry according to the scheduled list, so named as Automated Scheduled Telescope and also is the beginning of robotic telescope. Then Remotely Operated Telescope is proposed to meet the needs of observation, which could be controlled by remote users and observe astronomical objects automatically. Today observatory are highly integrated, more complex and more advanced. Telescope observatory not only needs automatic unattended operation, but also can be connected remotely. It can run without human's help and can adjust itself according to weather, equipment status and so on. This is the Robotic Autonomous Observatory, and some observatories have achieved this goal.

Robotic has two main important advantages, autonomous and remote. Autonomous means that a better use of telescope time by telescope's real-time follow-up, saving manpower in unattended operation mode, operation mistakes reduced, higher observation efficiency, and focusing science more but not operation logic. Remote means that users can be located anywhere to save time, share same telescope in different time to save cost and build observatory at high altitude and distant location to obtain the best observation environment. Robotic observatory has made significant effect in student education, for example the Bradford Robotic telescope, the original Micro-Observatory telescopes and so on(e.g.\ \citet{2017AstRv..13...28G},\ \citet{1996ASPC..101..380D}). These learning and observation experience could motivate students to continue higher level courses and scientific career. Even some high school students continued deeper research and produced papers(\ \citet{2011PASA...28...83F}). As a result of robotic telescope's huge advantages, more amateur astronomers have began to build their own observatory and have made many important scientific outputs, for example Xingming Observatory built in 2007 by an amateur astronomer. It is located in Xinjiang, China, and releases the figures by Popular Supernova Project.\footnote{Populsar Supernova Project reference web link \url{http://psp.china-vo.org/}.} This project platform is managed by Chinese Virtual Observatory, and any people can participant. Until now, 17 supernova and nova candidates have been reported, and 12 of them have been confirmed by optical spectrum. Public observatory and amateur astronomers, whether in education or scientific research, have been an important astronomy strength.

With economic progress, light pollution becomes worse and worse,\footnote{The light pollution map reference web link \url{https://www.lightpollutionmap.info}.} so that building observatory in remote areas becomes inevitable. There are some solutions by using modern robotic observatory mode, but usually too expensive and complex, and also not necessary to amateur astronomers. They usually organize their own system by themselves, especially for hardware equipment connection and related control. As a result of the limitation of technical condition and construction funds for amateur astronomers, their system usually consists of multiple sub-systems made by different people. This kind of combination is very rough and less compatible, so as to break down often. A stable remote unattended operation system suitable for amateur astronomer's observatory becomes very critical. In fact, remote unattended operation observatory is not a new conception, and it is the so called robotic telescope above, but a little difference and mainly faces the requirements of amateur astronomers' observatory. Here we propose a conception of engineering design for public observatory. It can not only satisfy observation requirement, but also save cost.

\section{The Conception of Engineering Design}
Nowadays public observatory also is not a single telescope, but an integration system with telescope, various sensors and so on. When design a common observatory hardware system, there are two key problems - the connection and intelligent control to equipment. We propose a hardware system as a bridge between observatory equipment and software system, and some criterions should be followed.

\begin{itemize}
\checklistitemize
\item Connect and control every equipment easily without any dependence on operation system and software platform.
\item Multiple connection modes to meet various users.
\item Hardware system itself should be clever to open or close some resource according to system status.
\end{itemize}

\articlefigure[width=.8\textwidth]{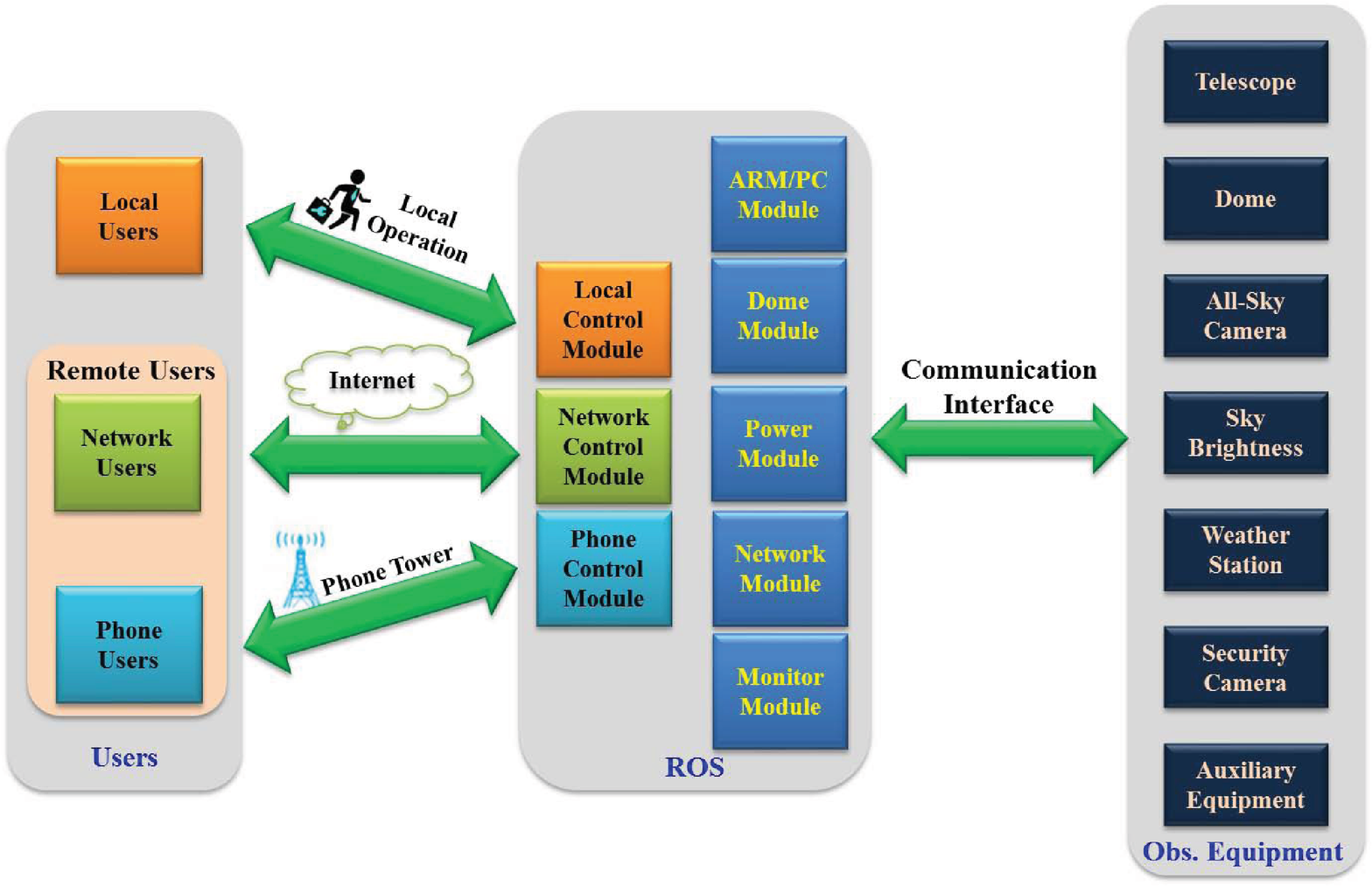}{P1-4_f1}{The framework for public observatory. \emph{Left:} Two kind of users - local users, remote users (including network users and phone users), are supported. \emph{Center:} The core conception of amateur observatory - ROS provides multiple connection modes to meet different users' needs and control logic to observation equipment. \emph{Right:} The necessary equipment in the observatory and also could be extended according to factual requirements. }

Based on these, we design a Remote Observatory System (ROS for short) to meet the criterions above, and make it to have the ability to connect and control the equipment in observatory. The framework for public observatory is showed in Figure \ref{P1-4_f1}. We define this system as a closed-loop system and make it has the capability to evaluate its operation through redundant inputs to detect errors. This system is made of multiple single chips, tiny internet chips and logic circuits, and could be reprogrammed. The ROS system consists of three control modules and five functional modules.

As a result of light pollution and actual demand, observatory could be located anywhere. Different control modules should be designed to meet different users. They are mainly used to transfer and analyse control commands. Three control modules are designed and listed next.
\begin{itemize}
\item Local Control Module - This module is the most basic part. Users can operate all the resource in the observatory, and it has the highest control priority.
\item Network Control Module - It has the lowest control priority and mainly used by remote users. Network protocol is independent of platform, and so it can be connected by any network equipment, for example computer, pad and so on.
\item Phone Control Module - This module is a special part and most useful for emergency control, for example network interruption. It connects each other by phone tower and the control is by message or voice.
\end{itemize}

The following is the five functional modules. Communication interface to observation equipment is mainly by internet or serial bus. They are used to connect and control observation equipment in observatory, for example telescope, dome, all-sky camera, sky brightness, weather station, security camera, other auxiliary equipment, etc.
\begin{itemize}
\item ARM/PC Module - This module is used to deploy related observation software for computing, data transfer, backup and telescope control, including equatorial mount, filter, focus and so on.
\item Dome Module - Dome open, follow-up and close.
\item Power Module - The power supply and control logic for every equipment.
\item Network Module - The network entrance and export. All network equipment should access and connect it.
\item Monitor Module - It not only monitor observatory equipment and its status, but also push status code to users and adjust resource's operation by predefined algorithm.
\end{itemize}

\section{Summary}
We propose a closed-loop hardware system as a bridge between observatory and users. It supports multiple control modules to meet different users, and provides internet and serial bus interface as the communication interface to connect observation equipment. The interface also could be extended according to factual requirements. It is a kind of open source hardware platform, and people could define control and transfer logic by themselves and then reprogramme it. Based on this hardware platform, we will develop software driver environment so as to access RTS2, ASCOM easily, and also make our own observation control software system special for public observatory in the future.

\acknowledgements This work is supported by National Natural Science Foundation of China (NSFC)(11503051, 61402325) and the Joint Research Fund in Astronomy (U1531111, U1531115, U1531246, U1731125, U1731243) under cooperative agreement between the NSFC and Chinese Academy of Sciences (CAS) and the Young Researcher Grant of National Astronomical Observatories, Chinese Academy of Sciences. We would like to thank the National R\&D Infrastructure and Facility Development Program of China, "Earth System Science Data Sharing Platform" and "Fundamental Science Data Sharing Platform" (DKA2017-12-02-XX). Data resources are supported by Chinese Astronomical Data Center (CAsDC) and Chinese Virtual Observatory (China-VO).

\bibliography{P1-4}

\end{document}